\documentclass[a4paper]{article}
\usepackage{amssymb}
\usepackage{amsmath}
\def\K{\rm KG}
\def\D{\rm Dirac}
\def\d{{\rm D}}
\author{M. Alimohammadi\footnote{alimohmd@ut.ac.ir} and  A. A. Baghjary
\\  {\small Department of Physics, University of Tehran,}
\\ {\small North Karegar Ave., Tehran, Iran.}}
\title{Klein-Gordon and Dirac particles in non-constant scalar-curvature background}
\date{}
\begin{document}
\maketitle
\begin{abstract}
The Klein-Gordon and Dirac equations are considered in a
semi-infinite lab ($x > 0$) in the presence of background metrics
$ds^2 =u^2(x) \eta_{\mu\nu} dx^\mu dx^\nu$ and
$ds^2=-dt^2+u^2(x)\eta_{ij}dx^i dx^j$ with $u(x)=e^{\pm gx}$.
These metrics have non-constant scalar-curvatures. Various aspects
of the solutions are studied. For the first metric with
$u(x)=e^{gx}$, it is shown that the spectrums are discrete, with
the ground state energy $E^2_{min}=p^2c^2 + g^2c^2\hbar^2$ for
spin-0 particles. For $u(x)=e^{-gx}$, the spectrums are found to
be continuous. For the second metric with $u(x)=e^{-gx}$, each
particle, depends on its transverse-momentum, can have continuous
or discrete spectrum. For Klein-Gordon particles, this threshold
transverse-momentum is $\sqrt{3}g/2$, while for Dirac particles it
is $g/2$. There is no solution for $u(x)=e^{gx}$ case. Some
geometrical properties of these metrics are also discussed.
\end{abstract}
\section{Introduction}
Studying the quantum mechanical effects of gravity is an important
and interesting branch of physics which has been started from the
early days of quantum mechanics. The simplest example of these
effects is the behavior of the nonrelativistic spinless quantum
particle, i.e. the Schrodinger equation, in the presence of
constant gravity \cite{1}. This phenomenon has been experimentally
verified by the famous experiment of Collela et al. \cite{2}. The
latest of these experiments is one reported by Nesvizhevsky et
al., in which the quantum energy levels of neutrons in the Earth's
gravitational field have been measured \cite{3,4}. Other aspects
of gravitational effects in quantum physics are appeared, for
example, in neutrino oscillation in gravitational background
\cite{5,6,7}, Berry phase of spin-$1/2$ particles moving in a
space-time with torsion \cite{8,9}, etc.

Another branch of researches in this area is the study of the
behaviors of Dirac and Klein-Gordon particles in the curved
background and distinguishes their physical characteristics. This
is an interesting subject since it makes clear the importance of
the spin of the particles, which is a purely quantum mechanical
property, in the gravitational interactions. Chandrasekhar, for
example, has considered the Dirac equation in a Kerr-geometry
background \cite{10}, with results which have been followed by
others \cite{11,12}.

A quick review in the literature of this field shows that the
number of integrable models is very few. For example in the case
of Schwarzschild metric, where the metric's components depend only
on one spatial coordinate $r$, the problem is too complicated to
be solved analytically. So trying to exactly solve some
relativistic quantum mechanical examples in curved background, may
shed light on this important topic and can help us to achieve more
insight into the realistic problems.

A class of background metrics which can be considered in this
area, is one which depends only on one spatial variable $x$. In
\cite{13}, a semi-infinite laboratory ($x>0$) has been considered
in background metric
\begin{equation}\label{1}
ds^2 = u^2(x)( - dt^2 + dx^2 ) + dy^2 +dz^2.
\end{equation}
In fact, an infinite barrier has been assumed in $x<0$ region. The
Klein-Gordon and Dirac equations have been studied for the
constant-gravity approximation of this background, i.e.
$u(x)\simeq 1+gx$, and some interesting features of this problem
have been discussed. For example it has been shown that in the
case of zero transverse-momentum, there exists an exact relation
between the squares of energy eigenvalues of spin-$1/2$ and spin-0
particles with same masses: $E_{\D}^2=E_{\K}^2+mg\hbar c$. So the
gravity clearly distinguishes between the Fermions and Bosons. The
scalar-curvature of metric (\ref{1}) is $R=0$.

In \cite{14}, another member of this class, that is the metric
\begin{equation}\label{2}
ds^2 =  - dt^2 + dx^2 + u^2(x)( dy^2 +dz^2),
\end{equation}
has been considered. For the case $u(x)=e^{-gx}$, the Klein-Gordon
and Dirac equations have been studied and their eigenfunctions
have been obtained. As an exact result, it has been shown that the
spin-0 particles have specific ground state energy :
$E_{\K}\geq\sqrt{m^2c^4+g^2c^2\hbar^2}$, while the spin-1/2
particles have the natural rest-mass energy ground state :
$E_{\D}\geq mc^2$. The scalar curvature of metric (\ref{2}) is
$R=6g^2$. So both metrics (\ref{1}) and (\ref{2}) have constant
scalar curvatures.

In this paper, we are going to study the behaviors of relativistic
spin-0 and spin-1/2 particles in one-variable-dependent-metrics
with non-constant scalar-curvature. For a semi-infinite lab with
an infinite potential barrier in $x<0$, we consider the following
two metrics:
\begin{equation}\label{3}
ds^2 = u^2(x)( - dt^2 + dx^2  + dy^2 +dz^2),
\end{equation}
\begin{equation}\label{4}
ds^2 =-dt^2 +u^2(x)( dx^2  + dy^2 +dz^2).
\end{equation}
It is worth nothing that in metrics (\ref{1}) and (\ref{2}), only
two of the metrics' components are nontrivial, while here we study
the next steps and take three ( in eq.(\ref{4})) and four ( in
eq.(\ref{3})) of the components nontrivial.

To make our problems more solvable, we assume $u(x)=e^{\pm gx}$,
which are the similar choices that have been considered in the
preceding cases. As the result, both metrics (\ref{3}) and
(\ref{4}) gain the $x$-dependent scalar curvatures, and the Dirac
and Klein-Gordon equations show interesting properties with
significant different behaviors. It is worth mentioning that
changing the variable from $x$ to $X$, defined by $X=\int u(x)dx$,
transforms the metric (\ref{4}) to (\ref{2}), but with new $u(x)$,
i.e. $u(x)\rightarrow f(X)$. Now if one decides to study both
metrics (\ref{2}) and (\ref{4}) with the same $u(x)$, as we do in
this paper, then these two metrics are independent.

The plan of the paper is as follows: In section 2, after fixing
our notations, we discuss the Klein-Gordon and Dirac equations in
background metric (\ref{3}) with $u(x)=e^{\pm gx}$. It is shown
that the spectrum of energy eigenvalues are discrete for
$u(x)=e^{gx}$ and continuous for $u(x)=e^{-gx}$. The discrete
spectrums are compared numerically and the geometrical properties
of the metrics, including their geodesics, are discussed. The
importance of geodesic in this problem is that it determines how
much it is possible to consider the $x=0$-plane as the floor of
the laboratory. Noting that the endpoints of all classical falling
particles are the floor, then only if the classical trajectories
finally intersect $x=0$, this plane can be considered as floor,
otherwise not. We see that this is the case for $u(x)=e^{-gx}$.

In section 3, the same is done for metric (\ref{4}) and it is
shown that only the case $u(x)=e^{-gx}$ is consistent with the
desired boundary conditions. For this $u$, it is shown that in
both cases, i.e. spin-0 and spin-1/2 particles, the spectrums have
interesting properties. They are continuous for $p<p_{0}$ and
discrete for $p>p_0$. $p$ is the transverse-momentum of the
particles and the value of $p_{0}$ is different for Dirac and
Klein-Gordon particles. The geometrical properties of metric
(\ref{4}) are also discussed. Finally in section 4, we review our
main results and bring some comments on metric
\begin{equation}\label{5}
ds^2 =u^2(x)(-dt^2+ dx^2)  +v^2(x)( dy^2 +dz^2),
\end{equation}
which is somehow a combination of two metrics (\ref{1}) and
(\ref{2}).
\section{Conformally-flat metric $ds^2 =u^2(x) \eta_{\mu\nu}
dx^\mu dx^\nu$}
 In a space-time with metric $g_{\mu\nu}$, the
Klein-Gordon equation in $c=\hbar=1$ unit is
\begin {equation} \label{6}
\left[\frac{1}{\sqrt{-\det g_{\mu\nu}}}\frac{\partial}{\partial
x^{\mu}}\left(\sqrt{-\det g_{\mu\nu}}
g^{\mu\nu}\frac{\partial}{\partial
x^\nu}\right)-m^2\right]\psi_{\K}=0.
\end{equation}
The Dirac equation in curved background is
\begin {equation}\label{7}
\left[\gamma^a\left(\partial_a +
\Gamma_a\right)-m\right]\psi_{\d}=0.
\end{equation}
$\gamma^a$s are the Dirac matrices and $\Gamma_a$s are spin
connections which can be obtained from tetrads $e^{a}$ through
\begin{align}\label{8}
de^a + \Gamma^{a}{}_b\wedge e^b=&0,\nonumber\\
\Gamma^a{}_b=&\Gamma^a{}_{cb}e^c,\nonumber\\
\Gamma_a=&-\frac{1}{8}\left[\gamma_b,\gamma_c\right]\Gamma^c{}_a{}^b.
\end{align}
The boundary conditions of the semi-infinite lab ($x>0$) with an
infinite potential barrier at $x=0$ are as follows. For
Schrodinger equation, the boundary condition is
$\lim_{x\rightarrow 0}\psi_{\rm sch}=0$, which comes from the fact
that the Schrodinger equation is second order in $x$, so
$\psi_{\rm sch}$ must be continuous at $x=0$. The same is true for
Klein-Gordon equation, so the same boundary condition exists:
\begin {equation}\label{9}
\psi_{\K}(0)=0.
\end{equation}
But the Dirac equation is of first order in $x$ and therefore
$\psi_{\d}$ can be discontinuous at $x=0$, if the potential goes
to infinity there. In this case, it can be shown that the desired
boundary condition is \cite{13}
\begin {equation}\label{10}
\left(\gamma^1-1\right)\psi_{\d}(0)=0.
\end{equation}
The square-integrability of wavefunctions $\psi_{\K}$ and
$\psi_{\d}$ in $(0,\infty)$ region also leads to
\begin {equation}\label{11}
\lim_{x\rightarrow \infty}\sqrt{\det
g_{\mu\nu}}\left|\psi\right|^2\rightarrow 0.
\end{equation}
\subsection{The Klein-Gordon equation}
For the metric (\ref{3}), the Klein-Gordon equation (\ref{6})
becomes
\begin{equation}\label{12}
\left(-\frac{\partial^2}{\partial t^2}+\frac{\partial ^2}{\partial
x^2}+ \frac{\partial^2}{\partial y^2}+ \frac{\partial^2}{\partial
z^2}+2\frac{u'}{u}
 \frac{\partial}{\partial x}-m^2u^2\right)\psi_{\K}=0.
\end{equation}
Since $u$ depends only on $x$, one may seek a solution whose
functional form is $\psi_{\K}
\left(t,x,y,z\right)=\exp\left(-iEt+ip_2y+ip_3z\right)\psi_{\K}(x)$.
Then $\psi_{\K}$ satisfies
\begin {equation}\label{13}
\left(E^2-p^{2}_{2}-p^{2}_{3}+ {\frac{d^2}{dx^2}}+2{\frac{ u'}{u}}
{\frac{d}{dx}}-m^2u^2\right)\psi_{\K}(x)=0.
\end{equation}
The above equation becomes solvable if we assume $u'/u$=constant.
So we take
\begin {equation}\label{14}
u(x)=e^{\pm gx}.
\end{equation}
Let us first consider $u(x)=e^{gx}$ case.

Substitute $u(x)=e^{gx}$ into eq.(\ref{13}), and defining
$\phi(x)$ through
\begin {equation}\label{15}
\psi_{\K}=e^{-gx}\phi(x),
\end{equation}
one finds
\begin {equation}\label{16}
\frac{d^2\phi\left(x\right)}{dx^2}+\left[g^2\left(\lambda^2-1\right)
-m^2e^{2gx}\right]\phi\left(x\right)=0,
\end{equation}
in which
\begin{align}\label{17}
p^2:=&p^2_2+p^2_3,\nonumber\\
\lambda^2:=&\frac{E^2-p^2}{g^2}.
\end{align}
In terms of new variable $X=(m/g)e^{gx}$, eq.(\ref{16}) reduces to
the modified Bessel equation
\begin {equation}\label{18}
X^2\frac{d^2\phi\left(X\right)}{dX^2}+X\frac{d\phi\left(X\right)}{dX}-
\left(X^2+1-\lambda^2\right)\phi\left(X\right)=0
\end{equation}
with solutions $K_\nu (X)$ and $I_\nu (X)$
$(\nu=\sqrt{1-\lambda^2})$. Using the asymptotic behaviors of
Bessel functions
\begin{equation}\label{19-1}
I_\nu(z)\sim z^\nu \ \ , \ \ K_\nu(z)\sim \frac{1}{z^\nu}\
(\nu\neq 0),
\end{equation}
in the limit $z\rightarrow 0$ ( with similar relations for Bessel
functions $J_\nu(z)$ and $Y_\nu(z)$, respectively), and
\begin{equation}\label{19-2}
I_\nu(z)\sim \frac{e^z}{\sqrt{z}} \ \ , \ \ K_\nu(z)\sim
\frac{e^{-z}}{\sqrt{z}},
\end{equation}
in the limit $z\rightarrow \infty $, it can be easily seen that
the boundary condition (\ref{11}) discards $I_\nu (X)$:
\begin{equation}\label{21-2}
\lim_{x\rightarrow \infty}\sqrt{\det
g_{\mu\nu}}\left|\psi_{\K}\right|^2 \sim e^{2(\nu +1)gx}
\rightarrow\infty.
\end{equation}
So the wavefunction becomes
\begin {equation}\label{19}
\psi_{\K}=Ce^{-gx}K_{\nu}(\frac{m}{g}e^{gx}).
\end{equation}
The boundary condition at $x=0$, eq.(\ref{9}), forces us to take
$\nu$ a pure imaginary number, since $K_{\nu}$ becomes zero only
when $\nu$ is pure imaginary $\cite{15}$. So one has to take
$\lambda^2>1$, which results the ground state energy of the spin-0
particles in background metric (\ref{3}) with $u(x)=e^{gx}$ to be:
\begin {equation}\label{20}
E^2\geq E^2_{min}=p^2c^2+g^2c^2\hbar^2.
\end{equation}
Other energy eigenvalues can be obtained by equation:
\begin {equation}\label{21}
K_{i\sqrt{\lambda^2-1}}(\frac{m}{g})=0,
\end{equation}
which clearly results in a  discrete spectrum.

In $u(x)=e^{-gx}$ case, a similar argument leads to
\begin {equation}\label{22}
\psi_{\K}=e^{gx}\left[C_1
I_\nu(\frac{m}{g}e^{-gx})+C_2K_\nu(\frac{m}{g}e^{-gx})\right],
\end{equation}
with $\nu=\sqrt{1-\lambda^2}$ and $\lambda$ defined through
eq.(\ref{17}). Since $\sqrt{\det g_{\mu\nu}}=e^{-4gx}$, the
function $I_\nu(\frac{m}{g}e^{-gx})$ satisfies the boundary
condition (\ref{11}) for all positive $\nu$s. But one has
$\sqrt{\det g_{\mu\nu}}\left|\psi_{\K}\right|^2\sim
e^{2(\nu-1)gx}$ for $K_\nu(\frac{m}{g}e^{-gx})$, which satisfies
(\ref{11}) only if $\Re(\nu)<1$. For $\lambda^2<1$,
$\nu=\sqrt{1-\lambda^2}<1$ is a real number and for $\lambda^2>1$,
$\nu$ is pure-imaginary number (so $\Re(\nu)=0$). Therefore in all
cases, one has $\Re(\nu)<1$ which implies that $C_1$ and $C_2$ in
eq.(\ref{22}) are arbitrary non-zero constants, up to the
normalization condition of wavefunction. The boundary condition
(\ref{9}) then gives
\begin {equation}\label{23}
C_1I_\nu(\frac{m}{g})+C_2K_\nu(\frac{m}{g})=0,
\end{equation}
which results in a continuous energy spectrum.
\subsection{The Dirac equation}
For metric (\ref{3}), the nonvanishing $\Gamma^a{}_b$s are
$\Gamma^0{}_1=\Gamma^1{}_0= (u'/u^2)e^0$,
$\Gamma^2{}_1=-\Gamma^1{}_2=(u'/u^2)e^2$ and $\Gamma^3{}_1
=-\Gamma^1{}_3=(u'/u^2)e^3$, from which
$\Gamma^0{}_0{}^1=-\Gamma^1{}_0{}^0=\Gamma^2
{}_2{}^1=-\Gamma^1{}_2{}^2=\Gamma^3{}_3{}^1=-\Gamma^1{}_3{}^3=u'/u^2$.
Therefore one finds $\Gamma_1=0$ and
$\Gamma_a=-(u'/2u^2)\gamma_1\gamma_a$ ($a=0,2,3$). Noting that
$\gamma^a\partial_a=\gamma^a e^\mu_a\partial_\mu$, in which
$e^\mu_a={\rm diag}(u^{-1},u^{-1},u^{-1},u^{-1})$, the Dirac
equation (\ref{7}) leads to:
\begin {equation}\label{24}
\left[\frac{1}{u}(\gamma^0\frac{\partial}{\partial t}+
\gamma^k\frac{\partial} {\partial x^ k})+\frac{3}{2}\frac{u'}
{u^2}\gamma^1-m\right]\psi_{\d}=0.
\end{equation}
Since $u=u(x)$, it is natural to take
$\psi_{\d}(t,x,y,z)=\exp(-iEt+ip_2y+ip_3z)\psi_{\d}(x)$. Defining
$\tilde{\psi}_{\d}(x)$ through
\begin{equation}\label{25}
\psi_\d(x)=u^{-3/2}\tilde{\psi}_\d(x),
\end{equation}
eq.(\ref{24}) then results in
\begin{equation}\label{26}
(O_1+O_2)\phi_\d(x)=0.
\end{equation}
In above equation, $O_1$, $O_2$ and $\phi_\d(x)$ are:
\begin{align}\label{27}
O_1=&-iE\gamma^1+\gamma^0\frac{d}{dx}-mu\gamma^1\gamma^0,\\
O_2=&i(p_2\gamma^2+p_3\gamma^3)\gamma^1\gamma^0,\label{28}\\
\phi_\d(x)=&\gamma^1\gamma^0\tilde{\psi}_\d(x).\label{29}
\end{align}
Using the fact that $\left[O_1,O_2\right]=0$, it may be possible
to choose the common eigenspinors for $O_1$ and $O_2$. The
eigenvalues of $O_2$ are $\pm ip$, with $p$ defined in
eq.(\ref{17}) and each eigenvalues are two-fold degenerate. For
$ip$-eigenvalue, the eigenspinors are
\begin{equation}\label{30}
\chi_1 = \left( \begin{matrix} i (p_2 - p)/p_3 \\ 1 \\
0 \\ 0 \end{matrix} \right) \ \ , \ \  \chi_2= \left(\begin
{matrix} 0
\\ 0 \\ i (p_2 + p)/p_3 \\ 1
\end{matrix} \right).
\end{equation}
So one can choose $\phi_\d(x)=\phi'_1(x)\chi_1+\phi_2(x)\chi_2$
and determines the unknown functions $\phi'_1(x)$ and $\phi_2(x)$
such that $\phi_\d(x)$ satisfies (\ref{26}). If we write $\phi_\d$
as
\begin{equation}\label{31}
\mathbf{\phi_\d}= \left( \begin{matrix}
\phi_1\\
\phi'_1\\
\phi'_2\\
\phi_2
\end{matrix} \right),
\end{equation}
then $\phi_1$ and $\phi'_2$ relate to  $\phi'_1$ and $\phi_2$ as
following
\begin{eqnarray}\label{32}
\phi_1=\frac{i(p_2-p)}{p_3}\phi'_1,\nonumber\\
\phi'_2=\frac{i(p_2+p)}{p_3}\phi_2.
\end{eqnarray}
Now it is sufficient to obtain two functions $\phi_1(x)$ and
$\phi_2(x)$, from which $\phi_\d(x)$ and therefore $\psi_\d(x)$
will be determined. Noting that $O_1\phi_\d=-ip\phi_\d$,
eqs.(\ref{27}) and (\ref{31}) result in:
\begin{eqnarray}\label{33}
\frac{d\phi_1}{dx}=p\phi_1-(E+mu)\phi_2,\nonumber\\
\frac{d\phi_2}{dx}=-p\phi_2+(E-mu)\phi_1.
\end{eqnarray}
It can be also easily shown that the boundary condition (\ref{10})
for $\psi_\d(0)$ reduces to the following boundary condition on
$\phi_1$ and $\phi_2$:
\begin{equation}\label{34}
(\phi_1(x) + \phi_2(x) )\mid_{x=0}=0.
\end{equation}
Defining $\psi_1$ and $\psi_2$ through:
\begin{eqnarray}\label{35}
\psi_1=\phi_1+\phi_2,\nonumber\\
\psi_2=\phi_1-\phi_2,
\end{eqnarray}
the differential equations (\ref{33}) then lead to:
\begin{eqnarray}\label{36}
\frac{d\psi_1}{dx}=\left(p+E\right)\psi_2-mu\psi_1,\nonumber\\
\frac{d\psi_2}{dx}=\left(p-E\right)\psi_1+mu\psi_2.
\end{eqnarray}

We first consider $u(x)=e^{gx}$ case. Introducing the new variable
$X=(2m/g)e^{gx}$, the differential equation of $\psi_1$ becomes
\begin {equation}\label{37}
X^2\frac{d^2\psi_1}{dX^2}+X\frac{d\psi_1}{dX}+\left(\lambda^2+
\frac{1}{2}X-\frac{1}{4}X^2\right)\psi_1=0,
\end{equation}
where $\lambda$ is defined in eq.(\ref{17}). Defining
$\tilde{\psi}_1$ through
\begin {equation}\label{38}
\psi_1=X^{-1/2}\tilde{\psi}_1,
\end{equation}
eq.(\ref{37}) leads to:
\begin {equation}\label{39}
\frac{d^2\tilde{\psi}_1}{dX^2}+\left(-\frac{1}{4}+\frac{1/2}{X}+
\frac{1/4+\lambda^2}{X^2}\right)\tilde{\psi}_1=0,
\end{equation}
which is Whittaker differential equation with solution
\begin {equation}\label{40}
\tilde{\psi}_1=e^{-X/2}X^{i\lambda+1/2}\left[C_1M\left(i\lambda,1+2i\lambda,X
\right)+C_2U\left(i\lambda, 1+2i\lambda,X\right)\right].
\end{equation}
$M\left(a,c,x\right)$ and $U\left(a,c,x\right)$ are confluent
hypergeometric functions. The boundary condition (\ref{11})
implies
$\psi_1(x\rightarrow\infty)=X^{-1/2}\tilde{\psi}_1(X)\mid_{X\rightarrow\infty}=0$.
But the asymptotic behavior of $M\left(a,c,x\right)$ is
$e^x/x^{c-a}$, so $C_1=0$. The second boundary condition
(\ref{34}) results in
$\psi_1(x=0)=X^{-1/2}\tilde{\psi}_1(X)\mid_{X=(2m/g)}=0$, which
leads to:
\begin{equation}\label{41}
\left(\frac{2m}{g}\right)^{i\lambda}U(i\lambda,1+2i\lambda,\frac{2m}{g})=0.
\end{equation}
This equation determines the discrete energy eigenvalues of Dirac
particles when they are in background metric (\ref{3}) with
$u(x)=e^{gx}$.

For $u(x)=e^{-gx}$, the same procedure leads to:
\begin {equation}\label{42}
\psi_1=e^{-Y/2}Y^{i\lambda}\left[C_1M\left(1+i\lambda,
1+2i\lambda,Y\right)+C_2U\left(1+i\lambda,
1+2i\lambda,Y\right)\right],
\end{equation}
where $\lambda$ is defined in eq.(\ref{17}) and
$Y\equiv(2m/g)e^{-gx}$. Here the boundary condition (\ref{11})
implies $\psi_1(Y\rightarrow 0) \rightarrow 0$ which can not
discard none of the constants $C_1$ and $C_2$. The energy
eigenvalues can be obtained by (\ref{34}), which results in
\begin {equation}\label{43}
C_1M\left(1+i\lambda,1+2i\lambda,2\frac{m}{g}\right)+C_2U\left(
1+i\lambda, 1+2i\lambda,2\frac{m}{g}\right)=0.
\end{equation}
Since $C_1$ and $C_2$ are arbitrary constants, up to the
normalization condition, eq.(\ref{43}) results in a continuous
energy spectrum.

It may be worth noting that wavefunctions of scalar and spin-1/2
particles in two background metrics which relate by a conformal
transformation, can be obtained from each other if the particles
are massless. If we consider the conformal transformation
$g_{\mu\nu}(x)\rightarrow {\bar g}_{\mu\nu}(x)=\Omega^2(x)
g_{\mu\nu}(x)$, then if $R=0$, one can show that
$\square\phi=0\rightarrow{\bar \square}{\bar \phi}=0$ in which
\cite{16}
\begin {equation}\label{44}
\bar{\phi}(x)=\Omega(x)^{(2-n)/2}\phi(x).
\end{equation}
$n$ is the dimension of space-time. For massless fermions, one
also has
\begin{equation}\label{45}
\psi(x)\rightarrow \bar{\psi}(x)=\Omega(x)^{(1-n)/2}\psi(x).
\end{equation}
In our problem, $R$ is zero ( for flat metric
$ds^2=\eta_{\mu\nu}dx^\mu dx^\nu$). Now if we take $m=0$ in
eq.(\ref{16}), the eqs.(\ref{15}) and (\ref{16}) result in
\begin {equation}\label{46}
\psi_{\K}(x)=u^{-1}\psi^{\rm flat}_{\K}(x),
\end{equation}
which is consistent with eq.(\ref{44}). Also if we put $m=0$ in
eq.(\ref{24}) and inserting eq.(\ref{25}) into eq.(\ref{24}), we
obtain
\begin {equation}\label{47}
\gamma^\mu\frac{\partial}{\partial x^\mu}{\tilde
\psi}_\d\left(x\right)=0,
\end{equation}
which is the free Dirac equation in flat space-time, i.e.
$\tilde{\psi}_\d=\psi^{\rm flat}_\d$. This shows $\psi^{\rm
non-flat}_\d=u^{-3/2}\psi^{\rm flat}_\d$ which is again consistent
with eq.(\ref{45}).

\subsection{Comparing the spectrums}
For $u(x)=e^{gx}$, the spectrums of spin-0 and spin-1/2 particles
can be obtained by eqs.(\ref{21}) and (\ref{41}), respectively.
None of these equations can be solved analytically and only in
$mc/g\hbar>>1$ limit, an approximate solution can be obtained for
eq.(\ref{21}) \cite{15}. The main difference between two spectrums
is that the Klein-Gordon eigenvalues has a ground state, i.e.
$\lambda^2_{\K}>1$, but for Dirac particle both $\lambda^2_{\D}<1$
and $\lambda^2_{\D}>1$ cases are possible. To obtain the numerical
values of energies ( $\lambda$s ), one must fix the values of $m$
and $g$ and then finds the roots of two equations (\ref{21}) and
(\ref{41}). For example for $mc/g\hbar=0.1$, Table 1 shows some of
the lowest energy levels of Dirac and Klein-Gordon particles. The
energies can be found by using eq.(\ref{17}):
$E=\sqrt{p^2c^2+\left(\lambda gc\hbar\right)^2}$.
\begin{table}[here]
\setlength{\tabcolsep}{.3pc} \caption{The lowest ten values of
$\lambda_{\K}$ and $\lambda_{\d}$ for $(mc/g\hbar)=0.1$}
\begin{center}
\begin{tabular}{|c|cccccccccc|}
\hline
$\lambda_{\K}$& 1.52&2.27&3.02&3.74&4.44&5.12&5.78&6.42&7.04&7.68\\
\hline
$\lambda_\d$& 0.85&1.82&2.65&3.42&4.14&4.84&5.51&6.17&6.82&7.46\\
\hline
\end{tabular}
\end{center}
\end{table}
\subsection{The metric properties}
The scalar curvature of metric (\ref{3}) is
\begin {equation}\label{48}
R=6u^{-3}u''.
\end{equation}
So for $u=e^{gx}$, $R=6g^2e^{-2gx}$ and for $u=e^{-gx}$ one has
$R=6g^2e^{2gx}$. Both $R$s are $x$-dependent, and in $u=e^{-gx}$
case, the scalar-curvature $R$ and the Kretschmann-invariant
$K=R_{\mu\nu\alpha\beta} R^{\mu\nu\alpha\beta}=6g^4e^{4gx}$
diverge at $x\rightarrow\infty$.

The classical trajectories of particles in these backgrounds are
also interesting. For $u=e^{\pm gx}$, one can show that
\begin{align}\label{49}
x\left(t\right)=&x\left(0\right)\mp\frac{1}{g}\ln\cosh\left[
\sqrt{1-v_\bot^2} g\left(t-t_0\right)\right],\nonumber\\
y\left(t\right)=&y\left(0\right)+v_{0y}t,\\
z\left(t\right)=&z\left(0\right)+v_{0z}t,\nonumber
\end{align}
in which $v_{0y}$ and $v_{0z}$ are arbitrary constants and $v_\bot
^2=v_{0y}^2+v_{0z}^2$. The $x$-component of velocity is
\begin {equation}\label{50}
v_x\left(t\right)=\mp\sqrt{1-v_\bot^2}\tanh\left[\sqrt{1-
v_\bot^2}g\left(t-t_0\right)\right].
\end{equation}
Note that in $\hbar=c\equiv 1$ unit, $v_x^2+v_\bot^2<1$.
Eq.(\ref{50}) shows that for $u=e^{gx}$, $v_x\left(t\right)$ is
always negative ( the particles fall in ($-x$)-direction ) and in
$u=e^{-gx}$ case, $v_x(t)>0$ and the particles fall in
($+x$)-direction ( toward the singular region ).
\section{The $ds^2=-dt^2+u^2(x)\eta_{ij}dx^idx^j$ metric}
In this section we study the Klein-Gordon and Dirac particles in
the presence of the metric (\ref{4}):
$$~~~~~~~~~~~~~~~~~~~~~~~~ds^2=-dt^2+u^2\left(x\right)\left(
dx^2+dy^2+dz^2\right)~~~~~~~~~~~~~~~~~~~~~~~~~~(4)$$.
\subsection{The spin-0 particles}
For metric (\ref{4}), the Klein-Gordon equation (\ref{6}) becomes
\begin {equation}\label{51}
\left[-\frac{\partial^2}{\partial
t^2}+\frac{1}{u^2}\left(\frac{\partial^2}{\partial x^2}+
\frac{\partial^2}{\partial y^2}+\frac{\partial^2}{\partial
z^2}\right)+\frac{u'} {u^3}\frac{\partial}{\partial
x}-m^2\right]\psi_{\K}=0.
\end{equation}
To make the above equation solvable, we assume that $1/u^2\propto
u'/u^3$ or $u(x)= e^{\pm gx}$. Let us first consider
$u(x)=e^{-gx}$ case.

Following the same steps of Sec.2.1, that is taking
$\psi_{\K}(t,{\mathbf r})= \exp(-iEt+ip_2y+ip_3z)\psi_{\K}(x)$,
defining $\psi_{\K}(x)=e^{gx}\phi(x)$ and $X=(k/g)e^{-gx}$, where
\begin {equation}\label{52}
k^2=E^2-m^2,
\end{equation}
eq.(\ref{51}) then reduces to
\begin{equation}\label{53}
\left(X^2\frac{d^2}{dX^2}+X^2-\eta_{\K}^2\right)\phi\left(X\right)=0,
\end{equation}
in which $\eta_{\K}=p/g$ $(p^2=p^2_2+p^2_3)$. The solutions of the
above differential equation are $\sqrt{X}J_\mu (X)$ and
$\sqrt{X}Y_\mu (X)$. $\mu$ is defined through
\begin {equation}\label{54}
 \mu=\sqrt{\eta^2+\frac{1}{4}}.
 \end{equation}
The limit $x\rightarrow\infty$ corresponds to $X\rightarrow 0$.
Since $\sqrt{\det g_{\mu\nu}}=e^{-3gx}$ and $Y_\mu (X)\sim
1/X^\mu$ (for $\mu\neq 0$), the boundary condition (\ref{11})
leads to $\sqrt{\det g_{\mu\nu}}\left|\psi_{\K}\right|^2 \sim
X^{2(1-\mu)}$. So for $\mu\leq\mu_0=1$, i.e. $p\leq
p_0=(\sqrt{3}/2)g$, both Bessel functions $J_\mu$ and $Y_\mu$ are
acceptable and the Klein-Gordon wavefunction in background metric
(\ref{4}) with $u=e^{-gx}$ becomes
\begin {equation}\label{55}
\psi_{\K}\left(x\right)=e^{(1/2)gx}\left[C_1J_\mu
\left(\frac{k}{g} e^{-gx}\right)+C_2Y_\mu
\left(\frac{k}{g}e^{-gx}\right)\right],~~~~~
\left(p\leq\frac{\sqrt{3}}{2}g\right).
\end{equation}
The eigenvalues are determined by eq.(\ref{9}) which results in a
continuous spectrum. For $\mu >1$, $Y_\mu (X)$ does not satisfy
the boundary condition (\ref{11}). Therefore the wavefunction is
\begin {equation}\label{56}
\psi_{\K}\left(x\right)=Ce^{(1/2)gx}J_\mu\left(\frac{k}{g}e^{-gx}\right),
~~~~~~~~~\left(p>\frac{\sqrt{3}}{2}g\right),
\end{equation}
and its corresponding eigenvalues can be obtained by
\begin {equation}\label{57}
J_\mu\left(\frac{k}{g}\right)=0,~~~~~~~\left(p>\frac{\sqrt{3}}{2}g\right).
\end{equation}

If we consider $u=e^{gx}$, instead of eq.(\ref{53}), one obtains
\begin {equation}\label{58}
\left(Y^2\frac{d^2}{dY^2}+Y^2-\eta_{\K}^2\right)\phi\left(Y\right)=0,
\end{equation}
where $\psi_{\K}\left(x\right)=e^{-gx}\phi\left(x\right)$ and
$Y\equiv(k/g)e^{gx}$. Again the solutions are
$\sqrt{Y}J_\mu\left(Y\right)$ and $\sqrt{Y}Y_\mu\left(Y\right)$,
but now the $x\rightarrow\infty$ limit corresponds to
$Y\rightarrow \infty$. Since the asymptotic behaviors of
$J_\mu(Y)$ and $Y_\mu(Y)$ in the limit $Y\rightarrow\infty$ are:
\begin {eqnarray}\label{59}
J_\mu\left(Y{}\right)\rightarrow\sqrt{\frac{2}{\pi
Y{}}}\cos\left(Y-
\mu \frac{\pi}{2}-\frac{\pi}{4}\right),\nonumber\\
Y_\mu\left(Y{}\right)\rightarrow\sqrt{\frac{2}{\pi
Y}}\sin\left(Y{}- \mu \frac{\pi}{2}-\frac{\pi}{4}\right),
\end{eqnarray}
and $\sqrt{\det g_{\mu\nu}}=e^{3gx}\sim Y^3$, the boundary
condition (\ref{11}) does not satisfy. So the Klein-Gordon
equation has no solution in this case.
\subsection{The spin-1/2 particles}
For Dirac particles in the presence of metric (\ref{4}), the
procedure is similar to one introduced in Sec.2.2, but with some
differences. Here, instead of eq.(\ref{25}), one has
$\psi_\d(x)=u^{-1}\tilde{\psi}_\d(x)$, and instead of (\ref{27}),
it is $O_1=-iEu\gamma^1+\gamma^0 d/dx-mu\gamma^1\gamma^0$, which
again $\left[O_1,O_2\right]=0$. Finally one arrives at, instead of
eq.(\ref{33}), the following equations for $\phi_1$ and $\phi_2$:
\begin{align}\label{60}
\frac{d\phi_1}{dx}=&p\phi_1-\left(E+m\right)u\phi_2,\nonumber\\
\frac{d\phi_2}{dx}=&-p\phi_2+\left(E-m\right)u\phi_1.
\end{align}
For $u=e^{-gx}$, by introducing $X=(k/g)e^{-gx}$, one finds
\begin {equation}\label{61}
\left(X^2\frac{d^2}{dX^2}+X^2-\eta_\d^2\right)\phi_1\left(X\right)=0.
\end{equation}
In this equation, $\eta_\d=\sqrt{p\left(p+g\right)}/g$, and $k$ is
defined through eq.(\ref{52}). It is interestingly seen that we
arrive at same equation for Klein-Gordon and Dirac particles, i.e.
eqs.(\ref{53}) and (\ref{61}), but with different $\eta$. Like the
pervious section, for $\mu\leq \mu_0=1$,
$\left(\mu=\sqrt{\eta_\d^2+1/4} \right)$, which now indicates
$p\leq p'_0=g/2$, both $\sqrt{X}J_\mu (X)$ and $\sqrt{X}Y_\mu (X)$
satisfy the boundary condition (\ref{11}), and therefore:
\begin {equation}\label{62}
\phi_1=e^{-(1/2)gx}\left[C_1J_\mu\left(\frac{k}{g}e^{-gx}\right)+
C_2Y_\mu\left(\frac{k}{g}e^{-gx}\right)\right],~~~~~~~\left(p\leq
\frac{g}{2}\right).
\end{equation}
The second function, $\phi_2$, can be obtained by the first
equation of (\ref{60}) and the resulting eigenvalues are
continuous.

For $\mu>1$, $Y_\mu (X)$ is not acceptable and therefore
\begin {equation}\label{63}
\phi_1=Ce^{-(1/2)gx}J_\mu\left(\frac{k}{g}e^{-gx}\right),~~~~~~~~~~~~~~~~~~~
\left(p>\frac{g}{2}\right).
\end{equation}
Using eqs.(\ref{63}) and (\ref{60}), the $x=0$ boundary condition,
i.e. eq.(\ref{34}), then leads to
\begin {equation}\label{64}
\left[1+\frac{2p+g}{2\left(E+m\right)}\right]J_\mu\left(\frac{k}{g}\right)+
\frac{k}{E+m}J'_\mu\left(\frac{k}{g}\right)=0,~~~~~~~~\left(p>\frac{g}{2}\right),
\end{equation}
where $J'_\mu$ is the derivative of $J_\mu$ with respect to its
argument. The above equation determines the discrete eigenvalues
of Dirac particle in the presence of metric (\ref{4}) with
$u=e^{-gx}$.

In the case $u=e^{gx}$, the result is same as in the Klein-Gordon
case and the $x\rightarrow \infty$ boundary condition can not be
satisfied.
\subsection{Spin-0 and spin-1/2's spectrums comparison}
The interesting feature of the spectrums of both equations, i.e.
the Klein-Gordon and Dirac equations, is that they are continuous
for low transverse-momentum $(p\leq p_0)$ and discrete for higher
one $(p>p_0)$. The value of $p_0$, however, differs for the two
cases, i.e. it is spin-dependent.
$(p_0)_{\K}=\left(\sqrt{3}/2\right)g$ and
$\left(p_0\right)_{\D}=g/2$. To find the energy eigenvalues, one
must fix the $m$, $g$ and $p$ values in eqs.(\ref{57})
and(\ref{64}) and then solves these equations numerically. For
example for $mc/g\hbar=0.1$ and $p/(g\hbar)=2$, the ten lowest
$k$-values for Klein-Gordon and Dirac particles are shown in Table
2. The energy eigenvalues can be obtained from this table by using
$E=\sqrt{k^2+m^2c^4}.$
\begin{table}[here]
\setlength{\tabcolsep}{.3pc} \caption{The lowest ten values of
$k_{\K}$ and $k_{\d}$ in $gc\hbar$ unit for $mc/g\hbar=0.1$ and
$p/(g\hbar)=2$}
\begin{center}
\begin{tabular}{|c|cccccccccc|}
\hline $k_{\K}$&0&5.21&8.50&11.71&14.88&18.05&21.21&24.36&27.51&30.66 \\
\hline $k_{\d}$&0&5.15&8.41&11.62&14.79&17.96&21.11&24.27&27.42&30.57  \\
\hline
\end{tabular}
\end{center}
\end{table}
\subsection{Classical characteristics of metric}
The scalar curvature of metric (\ref{4}) is
\begin {equation}\label{65}
R=4u^{-3}u''-2u^{-4}u'^2.
\end{equation}
So for $u=e^{-gx}$, we have $R=2g^2e^{2gx}$. Besides R, the
Kretschmann-invariant $K$ also diverges at $x\rightarrow \infty$,
$K=2g^4e^{4gx}$.

The equation of geodesics of this metric can be found as follows:
\begin{align}\label{66}
y=&k_1z+k_2,\nonumber\\
x\left(t\right)=&-\frac{1}{2g}\ln\left[e^{-2gx_0}+2g^2t^2\right],
\end{align}
where $k_1$, $k_2$ and $x_0$ are some constants. From the above
equation, $v_x$ is obtained as follows (in $c=1$ unit)
\begin {equation}\label{67}
v_x\left(t\right)=-\frac{2gt}{e^{-2gx_0}+2g^2t^2}
\end{equation}
In this space-time, the motion of classical particle is such that
the particle moves from $x=x_0$ toward $x=0$ with increasing speed
from $t=0$ to $t_\ast=e^{-2gx_0}/(\sqrt{2}g)$, and then its speed
decreases from $v_\ast=-(\sqrt{2}/2)e^{2gx_0}$ to zero at
$t\rightarrow \infty$.
\section{Conclusion}
In this paper, we consider the metrics which specified by a
one-variable function $u(x)$, specifically when $u(x)=e^{\pm gx}$,
and study the spin-0 and spin-1/2 particles in a semi-infinite
laboratory in the presence of these metrics.

First we consider the conformal metric $ds^2 =u^2(x) \eta_{\mu\nu}
dx^\mu dx^\nu$. For $u(x)=e^{gx}$, the Klein-Gordon and Dirac
eigenfunctions and the ground-state energy of spin-0 particles are
obtained exactly. The numerical values of discrete energy
eigenvalues can be calculated by using these eigenfunctions. For
$u(x)=e^{-gx}$, the exact eigenfunctions are obtained and it is
shown that the energy levels have continuous spectrum.

Second we study the metric $ds^2=-dt^2+u^2(x)\eta_{ij}dx^i dx^j$.
The problem has no solution for $u(x)=e^{gx}$. In the case
$u(x)=e^{-gx}$, the solutions depend on the transverse momentum
$p$. In all cases, we obtain the exact eigenfunctions. The
quantized values of $E_{\K}$ for $p>\sqrt{3}g/2$ and $E_{\d}$ for
$p>g/2$ can be obtained numerically.

In all the above mentioned cases, there are some detectable
differences between Fermions and Bosons, although there are some
analogy between them. One of the interesting observation is that
whenever the classical trajectory of particles crosses $x=0$, that
is the floor of the lab, then the quantum quantization appears in
the spectrum. There is, apparently, a correspondence between the
quantum-quantization and the falling property of the particles. In
the cases $u=e^{gx}$ of conformal metric and $u=e^{-gx}$ of the
second metric, where the quantization appears, the classical
trajectories are toward the ($-x$)-direction, while in $u=e^{-gx}$
case of the conformal metric, this is not the case.

Finally it may be interesting to bring some comments on metric
(\ref{5}). In this case, the Klein-Gordon wavefunction
$\psi_{\K}(x)$ satisfies
\begin{equation}\label{68}
 \left[-E^2+\frac{d^2}{dx^2}+2\frac{v'}{v}\frac{d}
{dx}-\frac{u^2}{v^2}\left(p_2^2+p_3^2\right)-m^2u^2\right]\psi_{\K}(x)=0,
\end{equation}
and the Dirac spinor's components satisfy
\begin {eqnarray}\label{69}
\frac{d\phi_1}{dx}=puv^{-1}\phi_1-\left(E+mu\right)\phi_2,\nonumber\\
\frac{d\phi_2}{dx}=-puv^{-1}\phi_2+\left(E-mu\right)\phi_1.
\end{eqnarray}
Here $\psi_\d=v^{-1}u^{-1/2}\tilde{\psi}_\d$, and other notations
are the same as ones introduced in Sec.2.2. By considering two
different exponential functions for $u(x)$ and $v(x)$, for example
$u(x)=e^{-ax}$ and $v(x)=e^{-bx}$, one may solve, not necessarily
analytically, eqs.(\ref{68}) and (\ref{69}). The scalar-curvature
$R$ of the metric (\ref{5}) for these specific $u$ and $v$ is
$R=6b^2e^{2ax}$. So if one chooses $a=0$ or $b=0$, then $R$
becomes constant and the metric (\ref{5}) reduces to metrics
(\ref{2}) and (\ref{1}), respectively.

{\bf Acknowledgement:}  We would like to thank the research
council of the University of Tehran for partial financial support.

\end{document}